\documentclass[conference]{IEEEtran}
\IEEEoverridecommandlockouts
\usepackage{cite}
\usepackage{amsmath,amssymb,amsfonts}
\usepackage{algorithmic}
\usepackage{graphicx}
\usepackage{textcomp}
\usepackage{xcolor}
\def\BibTeX{{\rm B\kern-.05em{\sc i\kern-.025em b}\kern-.08em
    T\kern-.1667em\lower.7ex\hbox{E}\kern-.125emX}}
\begin{document}

\title{Benefits and Limitations of Web3}

\author{\IEEEauthorblockN{Collin Connors}
\IEEEauthorblockA{\textit{Department of  Computer Science} \\
\textit{University of Miami}\\
Coral Gable, FL - 33124\\cdc104@miami.edu}
\and
\IEEEauthorblockN{Dilip Sarkar}
\IEEEauthorblockA{\textit{Department of  Computer Science} \\
\textit{University of Miami}\\
Coral Gable, FL - 33124\\sarkar@miami.edu}
}%

\maketitle

\begin{abstract}
Web3 provides users and service providers several benefits not found in Web2. However, despite the benefits provided, Web3 faces several obstacles that prevent the paradigm from gaining widespread adoption. Developers should understand the benefits and limitations of the technology in order to create more accessible Web3 smart applications. 
\end{abstract}


\section{Introduction}
\label{sec:Introduction}
Web3 combines older notions of a semantic web (Web3.0) with blockchain technology to allow for a more dynamic distributed web architecture. Using the Web3 architecture, end-users and service providers could see substantial benefits to data security, privacy, and overall user experience.

Our motivation for this work is to highlight for developers the benefits of Web3 while also emphasizing challenges the developers may face when implementing and utilizing a Web3 architecture in hopes of guiding developers in creating more accessible Web3 smart applications. We first discuss how a blockchain-based Web3 can vastly benefit all parties involved in creating and maintaining the World Wide Web. We then describe some obstacles that Web3 architectures currently face so that developers are aware of the limitations of the technology. We hope developers can use this knowledge to design their Web3 architectures and smart applications to be more accessible so that more end-users can reap the benefits of Web3. 

In the following sections, we will briefly discuss the history of the World Wide Web and how it has evolved since its inception \cite{Web_1.0_to_web3.0_2014} \cite{Web3_and_daos_2022}. We then discuss the semantic web (Web3.0), a precursor to the present Web3. We analyze the limitations of semantic web models and how blockchain has been proposed to address these challenges, leading to the current iteration of Web3. We highlight the benefits the Web3 paradigm can bring to both users and service providers. We then give an overview of the challenges Web3 architecture faces. Lastly, we conclude by providing guidance on overcoming the current limitations of the technology.

\section{Background}
\label{sec:Background}
\begin{figure*}
    \centering
    \includegraphics[width=.75\linewidth]{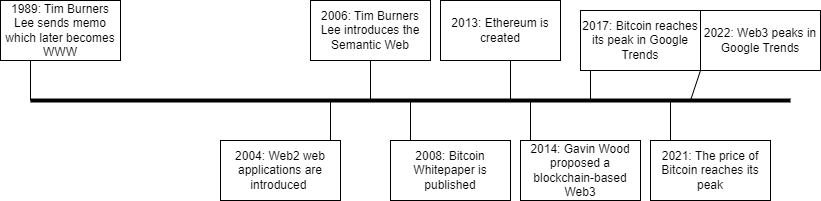}
    \caption{A timeline of key events in the history of the World Wide Web and the rise in popularity of Web3}
    \label{fig:timeline}
\end{figure*}

Since Web3 builds off previous generations of the World Wide Web, to best understand the advantages of Web3 it is critical to briefly review the history of the World Wide Web. Figure \ref{fig:timeline} provides a timeline of important events in Web technologies and blockchain technologies as they relate to Web3 \cite{Google_trends}. 

\subsection{Web1}
\label{sec:Web1}
In 1989, Tim Burners-Lee sent a proposal to management at CERN where he pointed out that the hierarchical system of keeping records that was in use made it difficult and time-consuming to search for related documents \cite{Burners-Lee}. To solve this problem, Burners-Lee suggested a new paradigm based on hypertext, which links records to each other through hyperlinks. Later, Burners-Lee combined hypertext with the internet to create the first version of the World Wide Web.  

This first version of the World Wide Web later became known as Web1. Features first introduced in Web1 have become ubiquitous with the modern web. The defining technology behind Web1 was Hypertext. Developers used HyperText Markup Language (HTML) to create content for web pages; web servers would then host the content created and allow end-users to find the content using Universal Record Locators (URLs); the end-users (will also be referred to as users) could utilize web browsers to access the web pages using the HyperText Transport Protocol (HTTP). While these technologies have been augmented and improved, they remain core to the modern World Wide Web. 

In Web1, web pages were static pages that utilized Hyperlinks to connect to other websites. These early web pages did not have the capability for end-users to generate new content. Instead, the early WWW acted as a directory where end-users could find HTML files from worldwide web servers. 

\begin{figure}
    \centering
    \includegraphics[width=.75\linewidth]{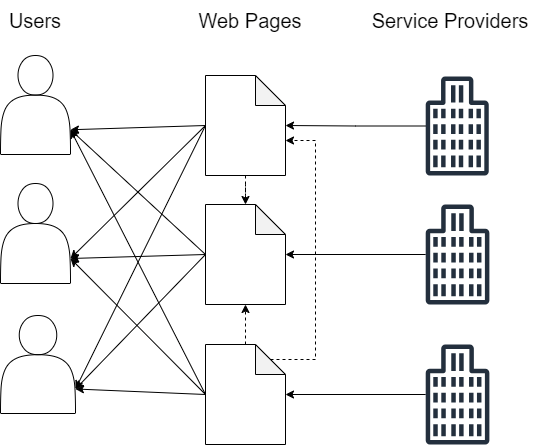}
    \caption{A high level overview of Web1's Read Only model.}
    \label{fig:Web1}
\end{figure}

This model of a Read-Only web provided end-users with an easy way to traverse the internet. Figure \ref{fig:Web1} shows a high-level view of Web1. In this paradigm, end-users can read data from web pages that are hyperlinked to each other. Various service providers host these web pages. Because there is no user interactivity, this era of the web was focused on companies providing content to end-users.

While the new technology developed to support Web1 was revolutionary, users quickly realized the limitations of a Read-Only web. This model requires content creators to host their web pages, leading to only those with specialized knowledge being able to create content for the World Wide Web. This meant that creating content for Web1, where businesses created most content, was highly inaccessible. Furthermore, the lack of communication between a content creator and the end-users consuming the content made it difficult for the end-users to engage with the content meaningfully. To address these limitations, Web2 was introduced to make the web more dynamic and interactive. 

\subsection{Web2}
\label{sec:Web2}
Web2 differentiates itself from Web1 by allowing for dynamic user-generated content. By introducing new technologies such as databases, Web2 allows for developing more complex web applications compared to the static web pages from Web1 \cite{Understanding_web_2.0_2007}. These dynamic applications allowed users to interact with content on the web in new ways.

\begin{figure}
    \centering
    \includegraphics[width=1\linewidth]{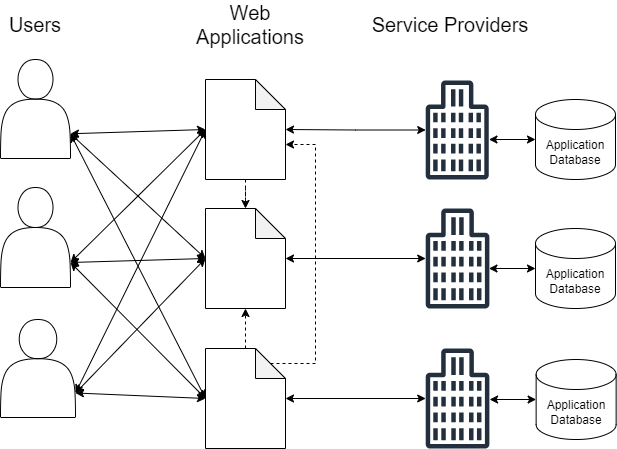}
    \caption{A high level overview of Web2's Read-Write data model}
    \label{fig:Web2}
\end{figure}

Figure \ref{fig:Web2} gives a high-level view of the Web2 model. This model builds on Web1, keeping ideas such as hypertext; however, it now allows users to store data in databases controlled by the web applications, allowing users to create their content and interact with the web on a deeper level. 

Because Web2 allows for interactivity, this paradigm is called the Read-Write web, where users can read the data just like in Web1 but can also write their data for web applications, a bidirectional data flow; users can send and receive data from applications. To handle the bidirectional data flow, more complex forms of storing data, such as XML and RSS, were created to support Web2. Web2 changes the focus of the web away from companies or site publishers and onto communities. With the rise of Web2, we begin to see new types of applications such as blogs, wikis, and social media. 

While Web2 improved the user experience over Web1 \cite{Web_1.0_to_web3.0_2014} \cite{Web3_and_daos_2022} \cite{Understanding_web_2.0_2007}, it came with new security challenges. Web applications were now storing users' data, putting these applications at risk of cyberattacks. This also decreased privacy as users now shared personal data with Web2 applications. Users also quickly noticed that the siloed nature of Web2 applications meant that data was often replicated across multiple applications, increasing a user's attack surface and making updating information across platforms inefficient. While many of these problems were identified in the early days of Web2, they still exist in modern Web2 applications. Researchers began exploring new models to make the web even more versatile and mitigate the limitations presented in the Web2 paradigm. 

\section{Web3}
\label{sec:Web3}
\subsection{What is Web3}
\label{sec:WhatIsWeb3}
As early as 2006, researchers noticed the limitations of Web2. To solve these problems, they began to propose a new paradigm of a semantic web\cite{The_emerging_web_of_Linked_Data_2009} \cite{The_dawn_of_semantic_search_2010}. The semantic web expands on Tim Burners-Lee's original ideas of linking documents to each other through hyperlinks, but rather than linking documents, individual pieces of data are linked to each other. 

To highlight the benefits of a semantic web, take an end-user of multiple Web2 social media applications. In each of these applications, the user has a short bio describing their interests. If the user finds a new interest, they must log in to each social media application and update each bio. This is time-consuming and inefficient since users must update the same data across multiple sites. The semantic web allows the user to update the data once and have that change reflected across all platforms on the web. 

This new paradigm reduced the data replication in Web2 and gave users more control over their data. However, Web2 applications are currently centralized and siloed off from one another, making it difficult to share data across platforms. Thus, implementing a large-scale semantic web requires decentralization.  

As cryptocurrency gained popularity and evolved, its underlying blockchain technology evolved in such a way that blockchain could provide a decentralized backbone for a large-scale semantic web \cite{Blockchain_for_decentralization_of_interne_2021} \cite{Blockchain-assisted_privacy-preserving_data_computing_architecture_2023}. Cryptocurrency, through the use of blockchain, allows for a trustless, decentralized exchange of data. The same principle required by semantic web, making blockchain a viable candidate to support the semantic web. 

Like the web, blockchain has evolved, adding more dynamic features \cite{JNCA_Survey}. Early blockchain projects such as Bitcoin were only focused on allowing users to transact cryptocurrencies. The next generation of blockchain, such as Ethereum, allowed users to create smart contracts, code that could be executed on a distributed state machine. With the dawn of generation 2 blockchains, developers began to combine semantic web ideas with decentralized smart contracts. Gavin Wood, one of the co-founders of Ethereum, first referred to this new paradigm as Web3 in 2014. 

\begin{figure}
    \centering
    \includegraphics[width=1\linewidth]{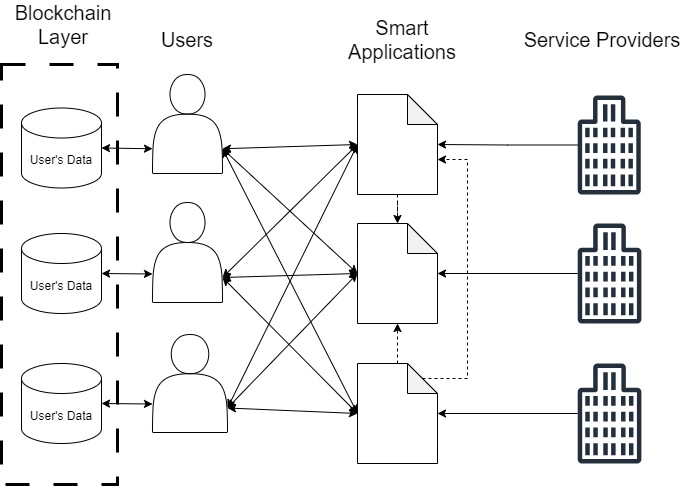}
    \caption{A high level overview of the Web3 Read-Write-Own data model.}
    \label{fig:Web3}
\end{figure}

In Web 2, users relied on various third-party service providers to store and maintain their data. For example, each platform must keep a copy of the user's bio in the social media use case. With Web3, users can store their bio on a public blockchain, such as Ethereum, via a smart contract, allowing each social media application to access this data. Thus, the user only needs to update their bio on the blockchain once and have that change reflected across all the social media platforms. In this Read-Write-Own model, the user controls their data and stores it on a blockchain, unlike Web2, where the platforms own the user's data and are responsible for storage. Figure \ref{fig:Web3}  highlights the Read-Write-Own model. 

Table \ref{tab:Compare} provides a summary comparison of the various generations of the web; more detailed comparisons can be found in \cite{Web_1.0_to_web3.0_2014} \cite{Web3_and_daos_2022} \cite{WEB3_2023}. 

\begin{table*}[]
\caption{Comparison of Web1, Web2, and Web3}
\resizebox{\textwidth}{!}{%
\begin{tabular}{|l|l|l|l|}
\hline
                      & \textbf{Web1}  & \textbf{Web2} & \textbf{Web3}   \\ \hline
\textbf{Data Model}   & Read           & Read Write    & Read Write Own  \\ \hline
\textbf{Key Technologies} & URL, HTTP, Web Server, Web Browser & Databases                        & Semantic Web, Blockchain \\ \hline
\textbf{Data Formats} & HTML           & XML, RSS      & RDF, RDFS, OWLS \\ \hline
\textbf{Content}          & Static Web Pages                   & Web Applications                 & Smart Applications         \\ \hline
\textbf{Context}      & Centralized    & Centralized   & Decentralized    \\ \hline
\textbf{Core Focus}   & Company        & Community     & Individual        \\ \hline
\textbf{Information Flow} & None                               & User-Application (Bidirectional) & Multi-directional        \\ \hline
\textbf{Data Owner}   & Site Publisher & Web Application Platform & Users           \\ \hline
\end{tabular}%
}
\label{tab:Compare}
\end{table*}

\subsection{Benefits of Web3}
\label{sec:BeefitsOfWeb3}

One of the immediate benefits of Web3's Read-Write-Own paradigm is that users are in control of their data. Thus, users can restrict who has access to their data and when access is granted. This is in contrast to Web2, where once users give their data to a platform, the platform is free to use that data in any manner, including selling the user's data to data brokers. 

Likewise, in Web2, if a user shares their data with an application, there is no guarantee that the user will be able to remove their data from the application. Users must trust that third-party application providers securely handle their data and will remove it when requested. In contrast, in Web3, the user controls the data and can grant or revoke an application's access to their data anytime. Furthermore, users can utilize blockchain to record which applications have access to their data and what data these applications have access to, allowing them to understand their attack surface and risk better. 

In Web3, users no longer need to trust third-party applications to keep their data secure. In Web2, a security-conscious user may review SOC audits or other security reports to ensure that an application is taking steps to keep data secure. However, in Web3, since the user owns their data, they can implement their controls to ensure that the data is secured. While this may be daunting for individual users, security-conscious businesses can greatly benefit from this paradigm. 

Another advantage of the Web3 paradigm is that users no longer need to replicate data across multiple applications. In Web2, a user may need to enter the same data into multiple applications, for example, updating their bio in multiple social media applications. With Web3, the user only needs to update this data one time. This makes interacting with multiple platforms on the web more streamlined, as data no longer needs to be replicated. For enterprises that need to manage complex relationships between applications, simplifying how data is entered can reduce costs and increase efficiencies.  

Likewise, since users no longer need to replicate their data in multiple places, this reduces a user's attack surface. In Web2, attackers targeting a user must find one weak application where a user has stored sensitive information to successfully steal that user's data. In Web3, the user only needs to worry about securing one data source.

Since applications no longer store data, the user and the provider's security risk is reduced. In Web2, an application accumulates data from many users, centralizing this data into one storage system, such as a SQL database. Since this storage system contains the data for many users of a centralized application, it makes it a high-value target for attackers. They only need to breach the storage system once to get the data for many users. In contrast, in Web3, the data is decentralized; thus, attackers no longer have a single point of attack, making it more difficult to breach all of the users' data. Application providers can reduce threat levels and liability by switching to Web3 and decentralizing users' data. It is critical to note that Web3 must be implemented correctly and securely for applications to experience a risk reduction; if implemented incorrectly, applications may expose themselves to higher levels of risk. 

Likewise, in a Web3 model, service providers no longer need to store users' data, which allows applications to spend less on storage. This cost reduction allows developers to focus their resources elsewhere to improve the application. In addition, since the storage is passed off to users, Web3 smart applications no longer need to worry about data storage when scaling the application. 

Overall, if implemented properly, Web3 can provide several benefits to both users and applications. These benefits include:
\begin{itemize}
	\item Giving users control of their data
	\item increasing users' privacy and security, 
	\item reducing data replication, 
	\item reducing the risk of cyberattacks on platforms, 
	\item and improving the developer experience. 
\end{itemize}

While Web3 offers many benefits, it has struggled to become mainstream due to several drawbacks of implementing the technology. 

\section{Limitations of Web3}
\label{sec:ChallengesOfWeb3}
Just as Web2's improvements to Web1 came with a new set of challenges, Web3 presents its own unique set of issues, including scalability, technological barriers, a desire for centralization, and an unwillingness to adopt the paradigm by large applications.  

A consistent challenge with blockchain technology has been the inability to scale effectively. Even more modern generation 3 blockchains fail to scale to the size necessary to support a global Web3 fully. However, for smaller enterprises, the web's modern blockchains can handle a more limited number of transactions, making Web3 valuable for businesses with local intranets. Web3 designers need to work to implement blockchain in a scalable manner to accommodate the entire web. Similarly, Web3 developers must know the scale limitations and design their applications to account for this. 

In addition to limited transaction throughput, blockchains often add cost per transaction. This additional cost adds an extra layer of decision-making to application developers. In some applications, the cost per transaction may offset or even eclipse the savings the developer experienced by reducing storage costs. Developers should carefully decide how to manage this cost to make more accessible Web3 applications, and developers should avoid passing this cost to users when possible. 

Web3 also proves difficult for non-technical users to utilize securely. In Web2, users trust that the application securely stores and maintains data. However, Web3 moves this burden to the users. Web2 development teams have security specialists who understand the risk to the user's data and have the resources to implement controls to mitigate risk. In Web3, the average user does not understand security to the depth necessary to secure sensitive data. This lack of knowledge may put users' data at risk, whereas a Web2 application would have provided some protection to the user. Since businesses have the resources to hire security experts, this challenge of Web3 can be overcome in an enterprise environment. Web3 developers should be aware of the security maturity of their users and design safeguards to guide users in utilizing smart applications securely. 

While technologists often hailed decentralization as a revolutionary idea, users can drift towards a centralized model in practice. In 2019, Raman et al. \cite{Challenges_in_the_decentralised_web_2019} examined the decentralized application Mastodon. Mastodon is a social media app like Twitter that relies on decentralized instances to host content. In this study, the researchers found that users, the infrastructure, and the content drove the application towards a more centralized model where only a few instances hosted most of the data—the pressure for applications to become centralized challenges developers trying to create a fully decentralized Web3. It is important to remember that currently most web users are only familiar with the centralization of Web2. Developers should understand this tendency of users to gravitate towards centralization and design systems to guide users towards a decentralized model. 

A final challenge to Web3 is the unwillingness of large applications to adopt a decentralized web. Tech giants such as Google or Amazon enjoy many benefits by being a large, centralized data source. They can aggregate and sell users' data and perform analytics to understand their user base better. It is unlikely that these companies will want to switch to Web3 since they will lose control of the data. For Web3 to become viable, users must pressure these companies into switching to a decentralized model. Because aggregating user data can drive profits, these centralized applications would only give up control of the user's data with extreme pressure from their user base.

While Web3 offers benefits over Web2, this model comes with its challenges. 
\begin{itemize}
	\item Since it relies on a blockchain, Web3 is difficult to scale and can be cost-inefficient. 
	\item Web3 can be challenging for non-technical users to implement and utilize securely. 
	\item Internal pressures from Web3 applications and their users can lead toward centralization. 
	\item The unwillingness of existing centralized applications to decentralize makes achieving a fully distributed Web3 challenging. 
\end{itemize}

While many of these limitations can be overcome in a business environment with proper resources, developers must work to make Web3 available to all users. 

\section{Conclusion}
\label{sec:Conclusion}
Since the dawn of the World Wide Web, developers have changed the architecture to improve the user experience. Developers continue to innovate, starting with simple web pages in Web1 and evolving to dynamic web applications in Web2 and then to smart applications in Web3. Each new generation of the web incrementally improves on the previous generation. However, as the web evolves, new challenges arise in each generation. While many of the challenges of Web2 have yet to be overcome, the benefits of Web2 greatly outweigh the limitations that have allowed Web2 to flourish. Currently, the impediments to Web3 have prevented it from gaining widespread adoption.

In future work, we plan to expand on our existing blockchain system \cite{Arxiv_PBL_System} to address the limitations that Web3 presents. If implemented carefully, Web3 can offer users substantial privacy benefits while reducing the burden of storing and maintaining users' data on service providers. 

With this work, we want to clarify Web3 by highlighting its history and evolution. This work outlines the benefits Web3 systems can provide users while showing the challenges developers face when implementing such a system. Web3's Read-Write-Own data paradigm is the next step in the evolution of the web, and while this model comes with its obstacles, developers can work to mitigate the risks and make Web3 more accessible.   

\bibliographystyle{plain}
\bibliography{ref}

\begin{IEEEbiography}[{\includegraphics[width=1in,height=1.6in,clip,keepaspectratio]{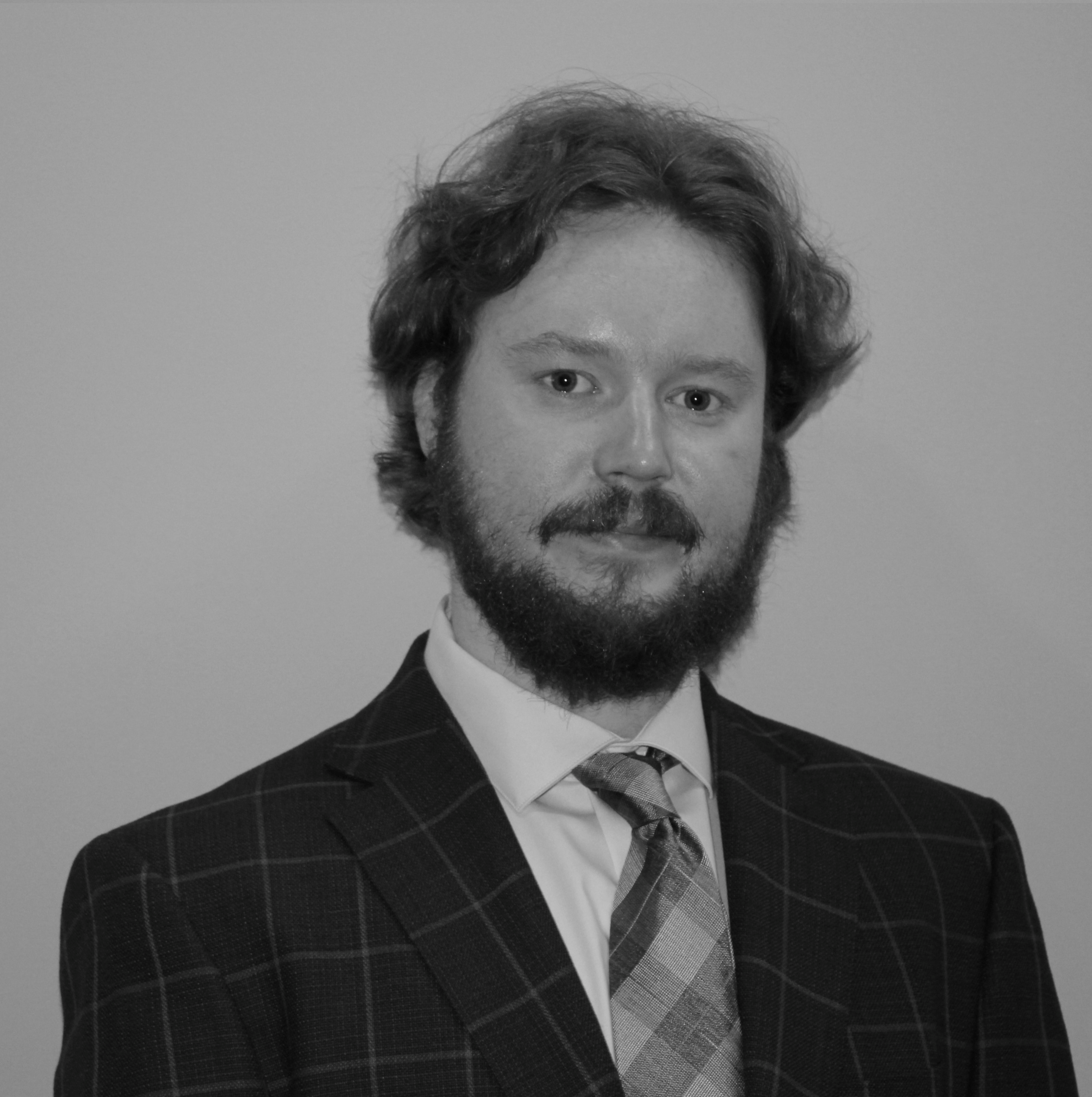}}]{Collin Connors}
received B.Sc. degrees in computer science and mathematics from the University of Miami, FL, in 2020. He is currently pursuing a Ph.D. degree with the Computer Science Department. His research interests include blockchain technologies, cybersecurity, and machine learning.
\end{IEEEbiography}

\begin{IEEEbiography}[{\includegraphics[width=1in,height=1.25in,clip,keepaspectratio]{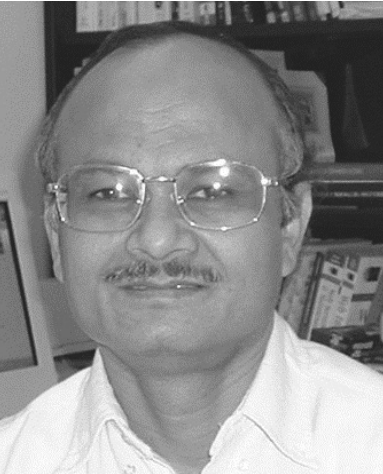}}]{Dilip Sarkar}
(SM’96) received the B.Tech.(Hons.) degree in electronics and electrical communication engineering from the Indian Institute ofTechnology at Kharagpur, Kharagpur, India, in 1983, the M.S. degree in computer science from the Indian Institute of Science, Bengaluru, India, in 1984, and the Ph.D. degree in computer science from the University of Central Florida, Orlando, in 1988. He is an Associate Professor of Computer Science at the University of Miami, Coral Gables, FL, USA. His research interests include concurrent transport protocols, parallel and distributed processing, neural networks, wireless sensor networks and their applications, and the security of virtual machines. In these areas, he has guided several theses and has authored numerous papers. He served as a Guest Editor of a Special Issue on Computer Communications on Concurrent Multipath Transport. For many years, he has served on ICC, Globecom, ICME, ICCCN, and INFOCOM program committees.
\end{IEEEbiography}

\end{document}